\documentclass[11pt,letterpaper]{article}
\usepackage{graphicx}
\usepackage{psfrag}
\usepackage{color}
\usepackage{amsbsy}
\usepackage{amssymb}    
\usepackage{multirow}   
\usepackage[square,numbers,sort&compress]{natbib} 
\def\dd{\mbox{d}}

\def\k{\kappa}

\def\ve{\varepsilon}

\def\r{{\bf r}}

\def\u{{\bf u}}

\begin{document}

\title{Mechanisms of receptor/coreceptor-mediated
entry of enveloped viruses}

\author{Sarah A. Nowak \\
Dept. of Biomathematics, \\
UCLA, Los Angeles, CA 90095-1766 \and 
Tom Chou\thanks{Corresponding author. Address: Department of Biomathematics, UCLA,
Los Angeles, CA, 90095-1766, U. S. A., Tel.: (310)-206-2787} \\
Dept. of Biomathematics, \\
Dept. of Mathematics, \\
UCLA, Los Angeles, CA 90095-1766}



\date{\today}

\pagestyle{myheadings}
\markright{Mechanisms of viral entry}

\maketitle

\abstract{Enveloped viruses enter host cells either through
endocytosis, or by direct fusion of the viral membrane envelope and
the membrane of the host cell.  However, some viruses, such as HIV-1,
HSV-1, and Epstein-Barr can enter a cell through either mechanism,
with the choice of pathway often a function of the ambient physical
chemical conditions, such as temperature and pH. We develop a
stochastic model that describes the entry process at the level of
binding of viral glycoprotein spikes to cell membrane receptors and
coreceptors.  In our model, receptors attach the cell membrane to the
viral membrane, while subsequent binding of coreceptors enables
fusion.  The model quantifies the competition between fusion and
endocytotic entry pathways.  Relative probabilities for each pathway
are computed numerically, as well as analytically in the high viral
spike density limit.  We delineate parameter regimes in which fusion
or endocytosis is dominant. These parameters are related to measurable
and potentially controllable quantities such as membrane bending
rigidity and receptor, coreceptor, and viral spike densities.
Experimental implications of our mechanistic hypotheses are proposed
and discussed.

}

\section{Introduction}

Entry mechanisms of enveloped viruses (viruses with a surrounding
outer lipid bilayer membrane) are usually classified as being either
endocytotic or fusogenic \cite{DIMITROV,MarshHelenius06}.  In fusion, the virus
membrane and the host cell membrane become joined by a pore.  Once the
two membranes are contiguous, the virus can directly enter the host
cell.  This process is typically mediated by binding of cell surface
receptors to glycoprotein spikes on the viral membrane surface, which
trigger embedded fusion peptides. In endocytosis, the host cell first
internalizes the virus particle, wrapping it in a vesicle before
either fusion with the endosomal membrane, or degradation of the virus
as the endosome is acidified.  Fusion of the endosomal membrane with
the viral envelope is often triggered by the acidic environment of the
endosome.

While viruses are typically thought to enter host cells via either
endocytosis or fusion, there is a growing list of viruses that are
known to enter cells though both pathways.  For example, influenza,
the avian leukosis virus (ALV), and Semliki Forest virus (SFV)
primarily enter cells via endocytosis followed by endosomal fusion
triggered by low pH. However, they have also been observed to directly
fuse with host cells if the pH of the extracellular environment is
lowered \cite{DuzgunesNir92,ANTIHARVARD,SFV}.  For some viruses ({\it
e.g.}, Influenza), the glycoprotein-receptor complexes that bind the
virus to the cell membrane initiate fusion under acidic conditions
encountered later in the process.  Many other viruses require the
binding of multiple cell surface receptors by multiple viral
glycoproteins for entry, and several such viruses have also been
observed to enter cells through their non-dominant pathway.

At least three of the twelve types of glycoproteins in the envelope of
the Herpes Simplex Virus-1 (HSV-1) bind cell surface receptors as
integral steps in viral entry.  As an initial step, glycoproteins gB
and gC bind to heparan sulfate (HS) proteoglycans on the cell surface,
attaching the virus to the host cell.  Once the viral and host cell
membranes are brought close to each other, glycoprotein gD can
associate with any of a number of cell receptors, including
Herpesvirus entry mediator (a tumor necrosis factor receptor),
nectin-1 (a member of the immunoglobulin superfamily), and
3-O-sulfated heparan sulphate (HS), to trigger fusion.  HSV-1 is known
to exploit at least three entry pathways: direct fusion with the host
cell membrane, endocytosis followed by fusion with an acidic endosome,
and endocytosis followed by fusion with a neutral endosome
\cite{ReskeKatz07}.
	
Epstein-Barr virus, another member of the Herpes virus family,
requires the binding of multiple glycoproteins to cell surface
receptors during entry.  When Epstein-Barr virions enter B-cells, the
glycoprotein complex gp350/220 binds to complement receptor type II
(CR2) to attach the virus to the host B-cell.  Fusion of the virus
with the cell membrane or endosome requires that glycoprotein gp42
associate with a HLA class II protein on the cell surface
\cite{HaanSpeck00}.  It is thought that the virus and cell membranes
must be brought close by gp350/220-CR2 binding before gp42 can bind a
HLA class II protein.  While the Epstein-Barr virus typically enters B
cells by endocytosis, eventually fusing with the endosome, it enters
epithelial cells by direct fusion with the plasma membrane.  There are
at least three models for the entry of Epstein-Barr virus into
epithelial cells.  1) An interaction between gp350/220 on the virus
and CR2 on the cell brings the membranes close.  Viral glycoprotein
complex gHgL can then interact with gHgL receptor on the cell,
triggering fusion.  2) The viral glycoprotein complex may directly
interact with its receptor on the cell membrane, triggering fusion.
3) The viral protein encoded by BMRF2 may interact with integrins on
the cell surface followed by gHgL-gHgLr binding \cite{HuttFletcher07}.

Human immunodeficiency virus (HIV) has also been shown to exploit both
entry mechanisms. HIV requires a receptor, CD4, for endocytosis, and
both CD4 and a coreceptor, usually CXCR4 or CCR5 to fuse with the host
cell membrane \cite{HIVENDO0,HIVENDO1}.  
The HIV coreceptor binds to the viral glycoprotein gp120 with a much
higher affinity if the glycoprotein spike is already bound to a CD4
receptor \cite{TrkolaMoore96, WuSodroski96, HillLittman97}. HIV
infects cells with which it fuses, and is typically inactivated upon
endocytosis \cite{HIVENDO1}.

A previous study \cite{CHOUBPJ} has examined the dynamics of viral
entry when a single type of cell receptor attaches the virus to the
cell membrane {\it and} induces fusion.  In this paper, we develop a
stochastic model that describes viral entry pathways in which binding
of a receptor to viral glycoprotein spikes is followed by binding of a
coreceptor to viral spikes.  In this model, the receptors are only
attachment factors and the coreceptors induce fusion.  The coreceptors
and receptors may both bind to the same viral glycoprotein, as is the
case for HIV, or they may bind to different glycoproteins or sets of
glycoproteins, as is the case for HSV-1 and the Epstein-Barr virus.
The selection of entry pathway is computed as a function of the
kinetic rates in the model.  We will discuss the sensitivity of
pathway selection to the local co-receptor-mediated fusion rate and
the rate of coreceptor binding.

Table 1 lists relevant physical parameters for HIV-1 and HSV that
guide assumptions of our model. Parameters values relevant to our
model, but not readily available, are left blank and await future
experimental investigation.

\section{Kinetic Model for Receptor-Coreceptor Engagement}
\label{MODEL}

\begin{figure*}[h]
\begin{center}
\includegraphics[width=6.4in]{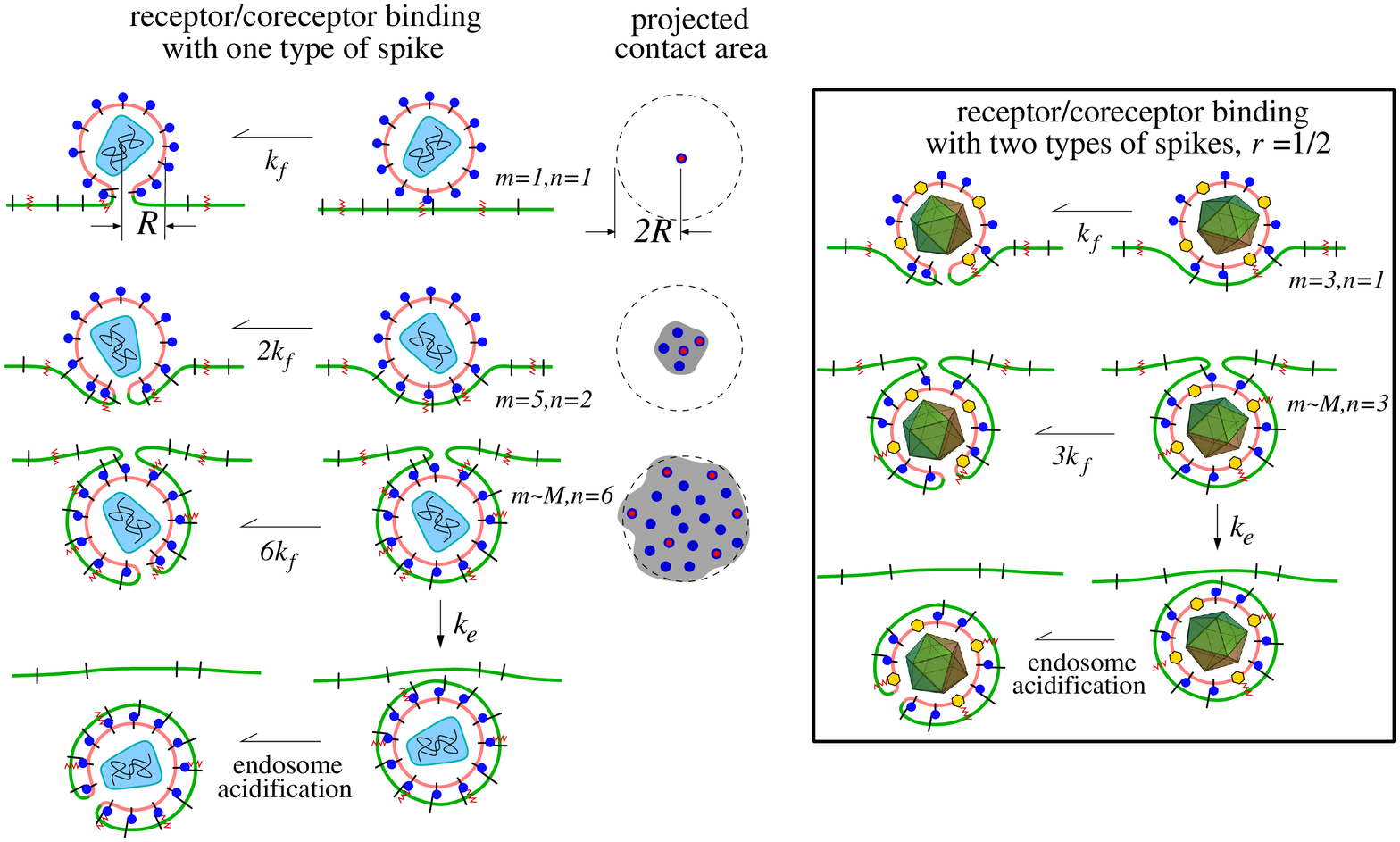}
\end{center} 
\vspace{-2mm}
\caption{}
\label{FIG1} 
\end{figure*}

Here, we derive a stochastic model describing the competition between
the endocytic and fusion viral entry pathways.  We assume that
receptors on a host cell membrane can bind to any one of $M$ spikes
uniformly distributed on the surface of a single virus, and that
coreceptors can bind to any one of $N$ spikes, which may be different
from the spikes to which receptors bind (see Fig.~\ref{FIG1}).
Receptor binding locally attaches the virus envelope to the cell
membrane, while coreceptor binding leads to the formation of
fusion-enabling complexes.  For simplicity, we consider the binding of
both receptors and coreceptors to be irreversible. Since binding
interactions between receptors and spikes can be very strong and/or
have low dissociation constants (see Table 1), this approximation is
consistent with physical parameters relevant to many viruses. However,
there is also evidence that the CD4-gp120 interaction is weak and can
dissociate during coreceptor recruitment \cite{WIRTZ}.

We assume that only those coreceptor-binding spikes in a region where spikes
are bound to receptors can bind coreceptors. This assumption is
appropriate if the receptors act as the attachment factor
that brings the viral and cell membranes close enough for the
coreceptor to bind.  For example, the binding of CR2 receptors to the
large gp350/220 glycoprotein complex on the Epstein-Barr virus 
typically precedes  attachment of fusion-inducing HLA class
II proteins to the smaller gp42 glycoprotein. 
This assumption also applies to HIV, since the affinity of coreceptors
for viral spikes increases significantly if the spike has already
bound a receptor \cite{TrkolaMoore96, WuSodroski96, HillLittman97}.

The ratio of coreceptor-binding spikes to receptor-binding spikes is
defined by $r = N/M$. For viruses where attachment receptors and
fusion initiating coreceptors attach to the same glycoprotein spike,
such as HIV-1, we can simply set $r=1$ in our model.  Although we
assume that each spike can bind at most only one receptor and/or one
coreceptor, experimentally inferred stoichiometries range from one to
a handful \cite{KABAT0,sodroski1,sodroski2}.  Our model can be
straightforwardly adapted to describe specific
receptor/coreceptor/spike stoichiometries.

%
%

In order for endocytosis to occur,  the virus must be fully wrapped by 
the cell membrane. 
We assume that when the virus is fully wrapped
all receptor-binding spikes
have a receptor attached. However, as more of the cell membrane
contacts the virus membrane through receptor binding, the rate of
binding of fusion-inducing coreceptors increases and fusion is increasingly likely.
It has been shown that for HIV, viral spikes act independently to
induce fusion \cite{YangSodroski05}, so we assume that the fusion rate
is proportional to the number of viral spike-receptor-coreceptor
complexes.  Although it is possible that spike-receptor-coreceptor
complexes that induce fusion do so in a cooperative manner, we are not
aware of any evidence that complexes aggregate in order to initiate a
cooperative response. From a modeling perspective, and in light of
experimental evidence \cite{YangSodroski05}, the most reasonable
assumption is that the randomly distributed spike-receptor-coreceptor
complexes locally induce irreversible fusion pore formation in an
independent and Poissonian manner. Thus, the total viral fusion rate
increases linearly with the number of spike-receptor-coreceptor
complexes formed.  Within our stochastic model, the likely pathway of
virus entry, endocytosis or fusion, will also depend on the specific
rates of receptor and coreceptor binding.

A mathematical framework representing our stochastic model is found by
considering, at any given time $t$, the probability $P_{m,n}(t)$ that
$m$ spikes are bound to a receptor and $n$ spikes are bound to a
coreceptor.  With the following definition of the relevant rates in
our problem

\vspace{4mm}

\noindent $\bullet$ $p_{m,n}$:  rate of binding an additional receptor

\vspace{2mm}

\noindent $\bullet$ $q_{m,n}$: rate of binding an additional coreceptor

\vspace{2mm}

\noindent $\bullet$ $k_f$: fusion rate for each spike-receptor-coreceptor complex

\vspace{2mm}

\noindent $\bullet$ $k_{e}$: rate of endocytosis (membrane pinch-off)
when all viral spikes are receptor-bound ($m=M$),

\vspace{4mm}

\noindent the probability $P_{m,n}(t)$ evolves according to the 
Master equation

\begin{equation}
\begin{array}{rlr}
\displaystyle {\partial P_{m,n}(t)\over \partial t} = &
p_{m-1,n}P_{m-1,n} +q_{m,n-1}P_{m,n-1} & 1\leq m \leq N-1, \, 1\leq n
\leq n^{*}-1 \\[13pt] \: & - (p_{m,n} + q_{m,n} +nk_{f})P_{m,n}, & n^{*}(m)\equiv \textrm{int}(rm)
\\[13pt] 
\displaystyle {\partial P_{m,n^{*}}(t)\over \partial t} = &
q_{m,n^{*}-1}P_{m,n^{*}-1}- (p_{m,n^{*}-1}+n^{*}k_{f})P_{m,n^{*}}, & \quad 1\leq m \leq M-1
\\[13pt] \displaystyle {\partial P_{m,0}(t)\over \partial t} = &
p_{m-1,0}P_{m-1,0}- (p_{m,0}+q_{m,0})P_{m,0}, & 1\leq m \leq M-1,
\\[13pt] \displaystyle {\partial P_{M,n}(t)\over \partial t} = &
p_{M-1,n}P_{M-1,n} +q_{M,n-1}P_{M,n-1} & 1\leq n \leq N-1,   \\[13pt]
\: & -(q_{M,n}+nk_{f}+k_{e})P_{M,n} & \: 
\end{array}
\label{eq:Master}
\end{equation}

\begin{figure}[ht]
\begin{center}
\includegraphics[width = 3.4in]{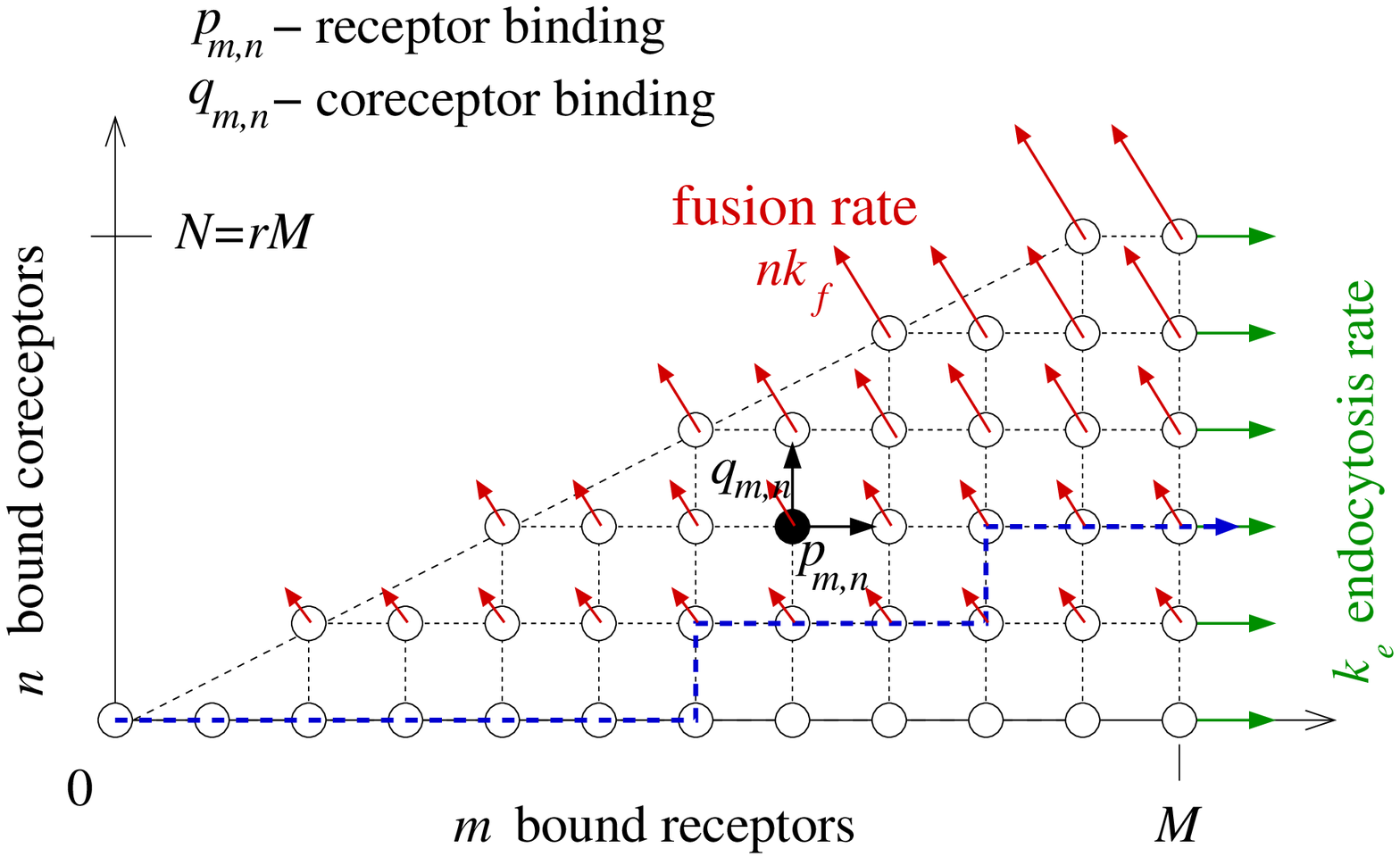}
\end{center}
\caption{}
\label{STATE}
\end{figure}

\noindent $\partial_{t}P_{0,0} =-p_{0,0}P_{0,0}$,
$\partial_{t}P_{M,0} = p_{M-1,0}P_{M-1,0}-(q_{M,0}+k_{e})P_{M,0}$, and
$\partial_{t}P_{M,N} = p_{M-1,N}P_{M-1,N}+q_{M,N-1}P_{M,N-1}-(Nk_{f}+k_{e})P_{M,N}$.  The
process depicted in Fig.~\ref{FIG1} and described by the above
equations can be represented by transitions within the $m,n$-state
space shown in Fig.~\ref{STATE}.

%
%

We treat all transitions in our model as Markovian, implicitly
assuming that they do not depend on past configurations.  This
assumption is appropriate if the attachment rates are kinetically
limited by membrane fluctuations or by receptor/coreceptor binding,
rather than by diffusion.  Diffusion-limited binding of receptors and
coreceptors gives rise to history-dependent kinetics and must be
treated within the framework of stochastic moving boundary
problems. Deterministic moving boundary problems relevant for virus
wrapping are treated in \cite{PAKWING} and \cite{GaoFreund05}. For
binding kinetics to not be diffusion-limited, receptors and
coreceptors must diffuse fast enough to replenish a receptor-depleted
region before the next binding event occurs. The time required for
concentration variations to diffuse away is $a_{r,c}/D_{r,c}$, where
$a_{r}$ and $a_{c}$ are the typical areas per receptor and coreceptor
on the cell surface, and $D_{r}$ and $D_{c}$ are their diffusion
coefficients in the cell membrane. Therefore, provided

\begin{equation}
p_{m,n} \ll D_{r}/a_{r} \quad \mbox{and} \quad q_{m,n} \ll D_{c}/a_{c},
\end{equation}

\noindent the history-independent binding assumption is justified. For
the HIV infection systems, CD4 receptor and CCR5 coreceptor
concentrations are approximately $10^{3}/\mu m^{2}$ and $60/\mu
m^{2}$, respectively.  Upon using the cell surface receptor and
coreceptor diffusion coefficients in Table 1, we find that $p_{m,n}
\ll 50/$s and $q_{m,n} \ll 3/$s are required for CD4 and CCR5
engagement to be kinetically limited, and not diffusion-limited.

\begin{figure}[ht]
\begin{center}
\includegraphics[width=3.3in]{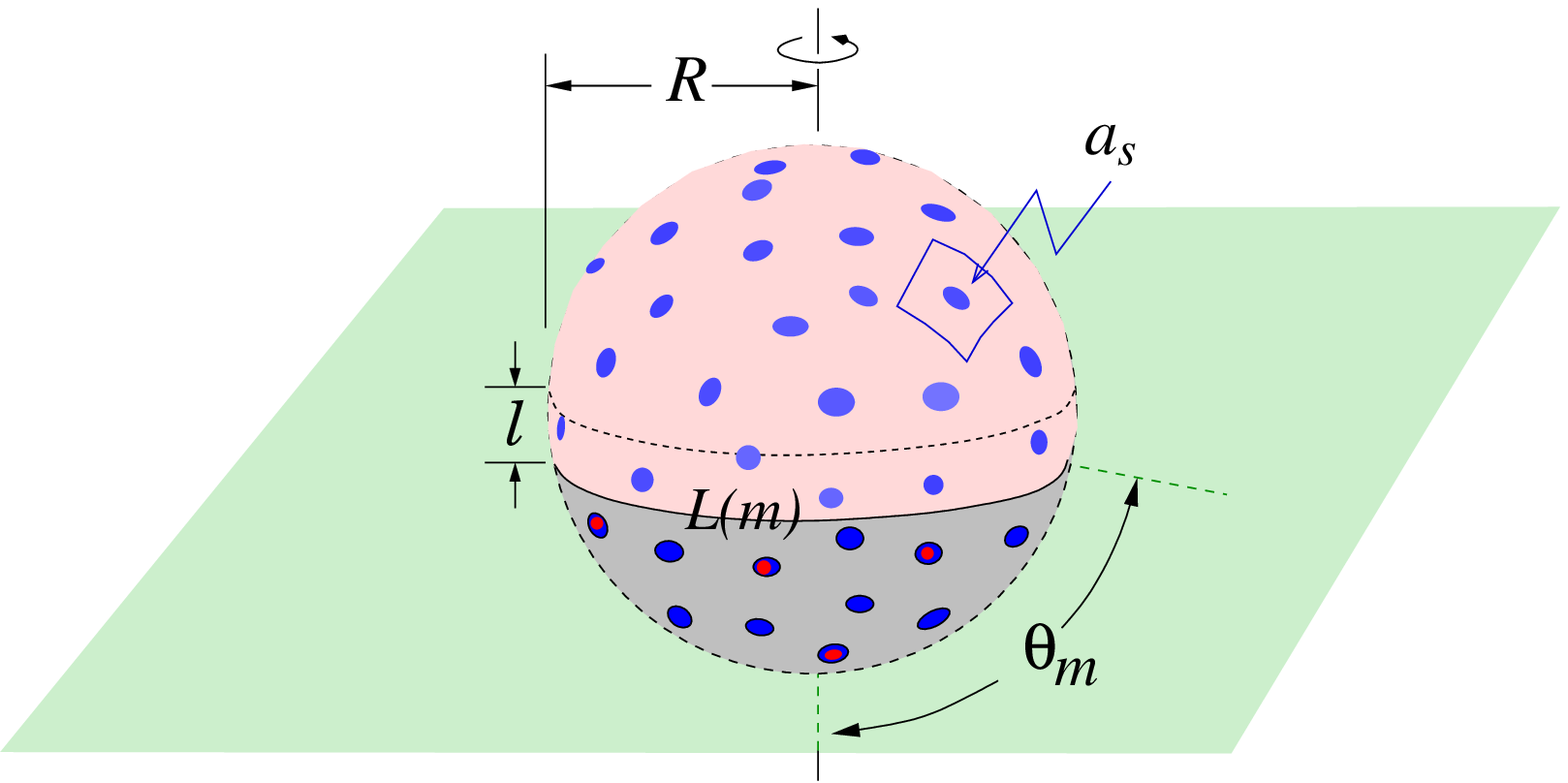}
\end{center} 
\vspace{-2mm}
\caption{}
\label{WRAP} 
\end{figure}

Although our main qualitative findings are independent of the precise
form for the attachment rates $p_{m,n}$ and $q_{m,n}$, we nonetheless
examine a specific physical model for these rates.  First, assume that
a receptor binds, with rate $p_{m,n}$, to only those spikes that are
within some small distance $\ell$ of the contact line $L(m)$ (see Fig.~\ref{WRAP}) where the
membrane detaches from the virus.  A functional form for this rate can
be derived by considering the number of ways additional receptors can
bind, given that there are already $m$ receptor-spike complexes making
up the contact region.  Fluctuations of the cell membrane will be
distributed in size with a typical scale $\ell$ (Fig. \ref{WRAP}). The
plasma membrane fluctuations, either thermally excited, or driven by
cellular processes such as cytoskeletal reorganization \cite{BUD}, can be caught
by the virus if they bring a receptor into the proximity of a
spike. As shown in Fig. \ref{WRAP}, the membrane wrapping process is a
Brownian ratchet that uses the spikes within a distance $\ell$ of the
contact line of length $L(m)$ to catch the cell membrane fluctuations.
%
%
The rate of attachment of an additional receptor can be written as
$p_{m,n} \sim \omega \ell L(m) a_{s}^{-1} a_{r}^{-1}$, where $\omega$
is an intrinsic attempt rate for binding and fluctuations of typical
size $\ell$, $a_{s}^{-1}$ is the viral spike concentration, and
$a_{r}^{-1}$ is the receptor concentration on the cell membrane.  The
term $\ell L(m) a_{s}^{-1}$ represents the probability that a membrane
fluctuation of typical size $\ell$ will encounter a spike on the viral
surface when $m$ receptors have already previously bound. The
approximate spherical geometry of this system gives $L(m) \approx
\left[1-(1-2m/M)^{2}\right]^{1/2}$, and since the area per spike is
$a_{s} \approx 4\pi R^{2}/M$, we find the coreceptor-independent,
receptor binding rate
 
\begin{equation}
\begin{array}{rl}
p_{m}(M) & \displaystyle \approx {\omega \ell M \over 4\pi R^{2}
a_{r}}\sqrt{1-\left(1-{2m\over M}\right)^{2}} \\[13pt] \: &
\displaystyle \equiv p_{1}M \sqrt{1-\left(1-{2m \over M}\right)^{2}},
\quad 1\leq m \leq M-1,
\label{PN}
\end{array}
\end{equation}


\noindent where $p_{1}M$ is the intrinsic rate of binding the second
receptor when initially one is bound. This intrinsic rate depends on a
number of physical parameters such as cell membrane bending rigidity
(through $\ell$ and $\omega$) and cell surface receptor concentration.
For stiff membranes under tension, a membrane wrapping a spherical
particle encounters an energy barrier near half-wrapping
\cite{Deserno-04}. This can be incorporated into the dynamics by
assuming $p_{1}$ has an $M$ dependence with a minimum near $m\approx
M/2$.  Other forms for $p_{m,n}$ can also be motivated \cite{LIGAND}
by considering the mechanics of wrapping \cite{Sun-06}.

The binding rate of coreceptors will be proportional to the integer
number of receptor-spike complexes that have not yet bound to
coreceptors:

\begin{equation}
q_{m,n} \approx q_{1,0} \textrm{int}(rm-n) \equiv q_{1}(n^{*}(m)-n),
\label{QN}
\end{equation}

\noindent where $q_{1}$ is the intrinsic rate of a coreceptor binding
to to a spike-receptor complex. 


Finally, we describe the fusion and endocytosis steps. The rates of
these processes, $k_{f}$ and $k_{e}$, are the least well measured. The
individual fusion rates $k_{f}$ depend not only on the particular
spike-receptor complex, but may also depend on other molecular factors
such as the lipid composition.
In model systems involving the gp41 fusion peptide of the HIV-1
glycoprotein-receptor complex, the fusion rate was found to be of the
order $k_{f} \sim 0.01$/s \cite{HaqueLentz02}. Physical models for
$k_{f}$ can also be motivated from phenomenological considerations of
fusion intermediates \cite{Siegel93,MARKIN,MONCK,KATSOV} and/or
estimated from computer simulations \cite{KATSOV,JAHN}.

The pinching-off of membrane vesicles in endocytosis is potentially a
more complex process activated by \textrm{GTP}ases such as dynamin
\cite{ROUX}.
The kinetics of this process may be akin to the ``kiss and run'' fast
mode of endocytosis at neuronal synapses. A wide range of rates (0.1/s
$< k_{e} < 20$/s) for synaptic vesicle kinetics has been reported
\cite{RoyleLagnado03}.

\section{Numerical Solution of Master Equation}

Solutions to Eq.~\ref{eq:Master} can be found numerically for up to
reasonably large values of $M$ and $N$.  From the resulting
probabilities, we construct time-independent quantities of interest.

The total time integrated probability $Q_{e}$ that the virus undergoes
endocytosis can be constructed from
\begin{equation}
Q_{e} = k_{e} \sum_{n=1}^{N} \int_0^{\infty}\!\! P_{M,n}(t)\dd t.
\label{eq:endoFlux}
\end{equation}
Similarly, the total time integrated probability $Q_f$ that the
virus undergoes fusion is
\begin{equation}
Q_f = k_{f}\!\!\!\sum_{m,n\leq  n^{*}}\!\!\!n\!\int_0^{\infty}\!\!P_{m,n}(t)\dd t = 1-Q_{e},
\label{eq:fusFlux}
\end{equation}

\noindent where the last equality arises from conservation of
probability and the assumption of non detaching receptors.
 
We solve for $Q_e$ and $Q_f$ by taking the Laplace transform of
Eq. \ref{eq:Master} and setting $s=0$.  If we define
$\tilde{P}_{m,n}=\int_{0^-}^{\infty} e^{-st} P_{m,n}(t) \dd t$, then
the endocytosis and fusion probabilities can be expressed as $Q_e
=\sum_n k_e \tilde{P}_{m,n}(s=0)$ and $Q_f = \sum_{m, n \leq n^{*}}
nk_f\tilde{P}_{m,n}(s=0)$, respectively.  We can also find the mean
times $\langle T_{e}\rangle$ to viral entry, conditioned upon
endocytosis, or $\langle T_{f}\rangle$, conditioned upon fusion:

\begin{equation}
\begin{array}{rl}
\langle T_{e} \rangle & \displaystyle =   
k_{e}Q_{e}^{-1}\sum_{n=1}^{N}\int_0^{\infty}\!\!\!t P_{M,n}(t)\dd t
\\[13pt]
\:  & \displaystyle  = -k_{e}Q_{e}^{-1}\sum_{n=1}^{N}
{\partial \tilde{P}_{M,n}(s=0)\over \partial s}
\label{eq:meanTime1}
\end{array}
\end{equation}
and
\begin{equation}
\begin{array}{rl}
\langle T_{f}\rangle & \displaystyle = k_{f}Q_{f}^{-1}\!\!\!\sum_{m,n< rm}
\!\!\!\!n\!\int_0^{\infty}\!\!\!tP_{m,n}(t)\dd t \\[13pt]
\: & \displaystyle =
-k_{f}Q_{f}^{-1}\!\!\!\sum_{m,n\leq n^{*}}\!\!\!\!n
{\partial \tilde{P}_{m,n}(s=0)\over \partial s}.
\label{eq:meanTime2}
\end{array}
\end{equation}

Finally, crucial to experimental considerations of
spike-receptor-coreceptor stoichiometry \cite{KABAT1}, we also compute
the mean numbers of receptors and coreceptors bound to the virus at
the moment of entry.  The mean number of receptors bound at the moment
of fusion is found from

\begin{equation}
\langle m_{f} \rangle = Q_{f}^{-1}\!\!\!\sum_{m,n \leq n^{*}}m
nk_f\tilde{P}_{m,n}(s=0).
\label{MF}
\end{equation}

\noindent The mean numbers of coreceptors bound at the moment of
fusion, and the mean number of coreceptors bound at the moment of
endocytosis (when all $N$ receptors are bound) can be similarly
obtained.

\section{Results and Analysis}

In this section, we discuss solutions of the Master Equation
(Eq.~\ref{eq:Master}) outlined in Section \ref{MODEL}.  For
simplicity, consider that the ratio of the number of spikes that
can bind coreceptors to the number of spikes that can bind receptors
is $r=1$ and that $M = N$.  This implies either that the number of
receptor-binding spikes equals the number of coreceptor binding
spikes, or that receptors and coreceptors both bind the same spikes,
as is the case for HIV.  The results for $r\neq 1$ are qualitatively
similar to the results of $r = 1$ when the replacement $k_f
\rightarrow rk_f$ is made (see Appendix).

\subsection{Pathway probabilities}

We first explore how the probability that the virus undergoes
endocytosis, $Q_e$, depends on problem parameters.  Since
$Q_{e}+Q_{f}=1$, it is sufficient to consider only $Q_e$. In
Fig.~\ref{fig:PeVkf}(a), $Q_e$ is plotted as a function of the
normalized fusion rate, $k_f/p_{1}$, for different values of the
normalized intrinsic coreceptor binding rate, $q_{1}/p_{1}$.  The
number of viral spikes, $M=N$, was chosen to be 100.
%
%
The probability that the virus undergoes endocytosis decreases with
increasing fusion rate, but a small coreceptor binding rate can
attenuate fusion even when $k_f$ is large.  In Figure
\ref{fig:PeVkf}(b), we plot the probability that the virus undergoes
endocytosis as a function of normalized fusion rate, $k_f/p_{1}$, for different values
of the normalized endocytosis rate, $k_e/p_{1}$.

%
%

The dependence of the endocytosis probability, $Q_e$, on the number of
viral spikes, $M$, is shown in Fig. \ref{fig:peVm}.  Although the
binding rate, $p_{m}$ increases with $M$ (see Eq. \ref{PN}), so do the
number of spikes that need to be engaged by receptors to achieve
the full wrapping required for endocytosis.  The time required to
fully wrap the virus is therefore constant with respect to $M$.
However, the fusion rate is proportional to the number of spikes with
coreceptors bound and is thus proportional to $N$.  As $N$
increases, the probability that the particle undergoes fusion before
it becomes fully wrapped increases, as illustrated by
Fig. \ref{fig:peVm}.  Figures \ref{fig:PeVkf} and \ref{fig:peVm}
clearly show a marked decrease in the endocytosis probability as the
fusion rate $k_{f}$ is increased.  

Since $k_{f}$ may vary greatly depending on physical chemical
conditions, as well as on viral species, it is important to estimate
the values of $k_{f}$ for which endocytosis or fusion is the dominant
mode of entry.
%
%
To better understand how $Q_e$ depends on $k_{f}$, we
consider the continuum limit of Eq. \ref{eq:Master}, appropriate for
large $M,N$. The probabilities of full wrapping and endocytosis, as
well as times to fusion and endocytosis, can be calculated
analytically by the method of characteristics (see Appendix).  Figure
\ref{fig:compare} compares our continuum limit analytic solution with
the exact numeric solution and agreement is good for $M,N\gtrsim 100$.
The analytic solution (see Eq. \ref{eq:ContinQe} in Appendix) provides
a guide for estimating the parameters for which endocytosis is likely.

Let us now dissect the entry dynamics and estimate values of $k_{e}$
and $k_{f}$ for which endocytosis will occur.  For certain parameters,
the virus is likely to fuse before it becomes fully wrapped.  In this
case, the probability that the virus reaches the fully wrapped state
will be small, and fusion will be the dominant mode of entry.  Only if
the virus is likely to become fully wrapped is endocytosis a possible
alternative to fusion.  Endocytosis will occur only if the probability
that $M$ receptors become bound to the virus, $P_M$, is order 1 {\it
and} endocytosis occurs more quickly than fusion once the virus
is fully wrapped.  For single receptor-spike complexes that attach
membranes {\it and} induce fusion \cite{CHOUBPJ}, previous asymptotic
analysis showed that

\begin{equation}
\left(\frac{k_f M^2}{p_{m}}\right) \ll 1
\label{OLDCOND}
\end{equation}
%
must be satisfied in order for the virus to become fully wrapped.  In
that analysis, $p_{m}$ was a typical receptor binding rate.

Analogous conditions for endocytosis can be found when both receptor
and coreceptor binding are required for fusion. These conditions can be
found numerically by computing $Q_{e}$ from Eqs. \ref{eq:Master} and
\ref{eq:endoFlux}.  However, upon using the specific forms for the
receptor and coreceptor binding rates given by Eqs. \ref{PN} and
\ref{QN}, the conditions can also be deduced from the wrapping
probability $P_{M}$ in the large $M$ limit.  From Eq. \ref{LNPTAU} in
the Appendix,

\begin{equation}
\ln{P_M}\approx -{rk_{f} M \over 2p_{1}}\left[{\pi\over 2}-\frac{1}{2\lambda}+
\frac{e^{-\lambda \pi}-\lambda^2}{2\lambda(\lambda^2+1)}\right],
\label{eq:Pfull}
\end{equation}
where $\lambda= q_1/(2p_1)$. This asymptotic expression 
allows us to determine when
the wrapping probability is appreciable.  If coreceptors are required
for fusion, as considered in this study, the expected behavior will be
similar to the single receptor model only if coreceptor binding is
faster than receptor binding.  Indeed, when $q_{1}\gtrsim p_{1}$
($\lambda \gtrsim 1$), we find that the condition
\begin{equation}
{r k_f M\over p_{1}} \ll 1
\label{eq:fastCondition}
\end{equation}
is required for full wrapping.  Since $p_{m} \sim p_{1}M$, we recover
the condition (Eq. \ref{OLDCOND} here)  given in \cite{CHOUBPJ} when $r=1$.
Figure \ref{fig:nonDim}(a) shows the numerically computed probability
of full wrapping, $P_M$, as a function of $r k_f M/p_{1}$ (with
$r=1$).  The condition for full wrapping given by
Eq. \ref{eq:fastCondition} holds even when parameters are individually
varied over a wide range of values.

%
%

Now consider the condition for full wrapping when coreceptor binding
is slow compared to receptor binding. For extremely small $q_{1}/p_{1}
= 2\lambda \ll 1/N$, the $M, N\rightarrow \infty$ continuum limit for
$P_{M}$ (Eq.~\ref{eq:Pfull}) is not appropriate. 
When coreceptor binding is extremely small, no coreceptors bind, and
the virus always becomes fully wrapped independent of
$k_{f}$. However, for $1/N \ll q_{1}/p_{1} \ll 1$, (or $1/N \ll
\lambda \ll 1$). The condition for $P_M \sim 1$ derived from
Eq. \ref{eq:Pfull} is

\begin{equation}
\left({r k_{f}M \over p_{1}}\right)\left({q_{1}\over p_{1}}\right) \ll 1.
\label{eq:slowCondition}
\end{equation}
If condition \ref{eq:fastCondition} is satisfied, condition
\ref{eq:slowCondition} will be satisfied provided that $q_{1} \leq
p_{1}$.  Thus, condition \ref{eq:fastCondition} is sufficient for the
virus to become fully wrapped; however, because slow coreceptor
binding can limit the effects of a fast fusion rate, condition
\ref{eq:fastCondition} is not necessary, particularly when coreceptor
binding is slow. In other words, even if $r k_{f}M/p_{1}$ is large, as
long as $q_{1}/p_{1}$ is small enough, condition
\ref{eq:slowCondition} can still be satisfied and full wrapping can
still occur. In Fig. \ref{fig:nonDim}(b), $P_M$ is plotted as a
function of $(r k_f M/p_{1})(q_{1}/p_{1})$ with various parameters
independently varied.  When coreceptor binding is slow, the condition
given by Eq. \ref{eq:slowCondition} is found to predict whether the
virus is likely to reach the fully wrapped state. Although we have
used the particular binding rates $p_{m}$ and $q_{m,n}$ from
Eqs.~\ref{PN} and \ref{QN}, analogous conditions for $P_{M} \sim 1$
can be motivated for general binding rates (see Appendix).

We can now derive sufficient conditions for endocytosis
after the virus becomes fully wrapped.  In the case where the
coreceptor binding rate is large compared to the receptor binding
rate, we expect that when the virus reaches the fully wrapped state,
nearly all $N$ spikes will be coreceptor-bound. Once the virus is
fully wrapped, it fuses with the cell membrane with total (and
maximal) rate $Nk_{f}$, while it is endocytosed by the cell with rate
$k_{e}$. Thus, endocytosis will be the dominant mode of viral entry
if

\begin{equation}
k_e \gg k_f N = rk_{f}M.
\label{eq:simpleKe}
\end{equation}
Provided the virus has a high probability of reaching the fully
wrapped state, $k_e \gg N k_f$ is always a sufficient, but not always
a necessary condition for endocytosis.  When coreceptor binding is not
fast, we will typically need to consider the full solution given by
Eq. \ref{eq:ContinQe} in order to determine when endocytosis is
likely. However, we can consider the limiting case where the
coreceptor binding rate is small compared to {\it both} the receptor
binding rate ($q_{1} \ll p_{1}$), and the fusion rate ($q_{1}N \ll
k_{f}$).  In this case, we can assume that fusion is limited by the
coreceptor binding rate, and the condition required for efficient
endocytosis is
\begin{equation}
k_e \gg q_{1} N = rq_{1}M.
\label{eq:otherKe}
\end{equation}
The conditions described above for efficient endocytosis are
summarized in the Discussion and Conclusions section and delineated in
a parameter-space ``phase diagram.''

\subsection{Mean entry times}

We now investigate $\langle T_{e} \rangle$, the mean viral entry time
via the endocytosis pathway, and $\langle T_{f}\rangle$, the mean
entry time via the fusion pathway. The normalized mean times are
computed from Eqs. \ref{eq:meanTime1} and \ref{eq:meanTime2} and are
plotted in Fig. \ref{fig:meanTime} as a function of the fusion rate
per bound coreceptor, $k_{f}$.  Endocytosis is governed by two
potentially rate-limiting steps: viral wrapping, and the final
endocytosis step (pinching-off of the cell membrane).  For small
$k_e$, the endocytosis step is rate limiting, and $\langle T_{e}
\rangle$ scales as $1/k_e$ when $Q_e \approx 1$.  For the parameters
used in Fig. \ref{fig:meanTime}(a), the receptor binding rates are
much faster than the endocytosis rate; thus, $k_e$ is the limiting
rate constant.  As the fusion rate $k_{f}$ increases, both the
probability of endocytosis, $Q_{e}$, and the mean endocytosis times,
$\langle T_{e}\rangle$, decrease.  One might initially expect $\langle
T_f \rangle$, but not $\langle T_e\rangle$ to decrease with increasing
$k_f$.  However, we expect there to be some distribution of times at
which the virus becomes fully wrapped.  A larger fusion rate will
preferentially annihilate trajectories that take longer to reach the
fully wrapped state.  Therefore, only trajectories that quickly reach
the fully wrapped state survive to $m=M$ and participate in
endocytosis, resulting in a decreased $\langle T_{e}\rangle$ when
$k_{f}$ is increased.

%
%

In Fig. \ref{fig:meanTime}(b), we plot the normalized mean entry times
as a function of $k_e/p_{1}$, the normalized endocytosis rate.  We
find that as we increase $k_e$, the mean time $\langle T_{e}\rangle$
decreases and then plateaus.  The plateau occurs when $k_e$ is
sufficiently fast that endocytosis is no longer rate-limiting.
Rather, membrane wrapping is the rate limiting step, and $\langle
T_{e}\rangle$ becomes independent of $k_e$.

%
%

\subsection{Mean receptor/coreceptors bound at entry}

Finally, consider the mean numbers of receptors and coreceptors bound
to the virus at the time of entry.  In Fig. \ref{fig:meanRecept}(a),
we plot $\langle m_{f} \rangle$, the mean number of receptors bound
when fusion occurs, and $\langle m \rangle$, the mean number of
receptors bound when the virus enters the cell through either pathway,
as functions of $q_{1}$, the coreceptor binding rate.  As $q_{1}$
increases, $Q_e$ decreases, and the virus is more likely to fuse with
the host cell.  Because the virus fuses more rapidly, there is less
time for receptors to bind and $\langle m \rangle$ decreases.
Fig. \ref{fig:meanRecept}(b) shows the mean number of coreceptors
bound to the virus at the time of entry.  For very small coreceptor
binding rates, the virus typically undergoes endocytosis before a
coreceptor can bind, and $\langle n_{e} \rangle \ll 1$, where $\langle
n_{e} \rangle$ is the mean number of coreceptors bound when the virus
undergoes endocytosis. However, at least one coreceptor must bind for
fusion to occur; therefore, when $q_{1}$ is small, the conditional
mean number of bound coreceptors $\langle n_{f} \rangle \approx 1$.
As $q_{1}$ becomes large, the probability that the virus undergoes
endocytosis becomes small, but the mean number of coreceptors bound to
the viruses that do undergo endocytosis approaches $N = 100$.  We know
that when $q_{1}$ is large, $n \approx rm$.  Since full wrapping ($m =
M$) is required for endocytosis to occur, we also expect $\langle
n_{e}\rangle \approx N$.

\section{Discussion and Conclusions}

We have developed a stochastic model describing the binding of
receptors and coreceptors to viral glycoprotein spikes, and the
subsequent competition between endocytosis and fusion during entry of
enveloped viruses.  Receptors function as simple attachment factors in
our model, while subsequent binding of coreceptors enables fusion. We
found parameter regimes in which endocytosis is favored and derived
analytic expressions for the probability of endocytosis in the large
spike number limit ($M, N\rightarrow \infty$).  Since the endocytosis
and fusion rates, $k_{e}$ and $k_{f}$, are difficult to measure, we
summarize our results by a ($k_{f}-k_{e}$) ``phase diagram'' defined
by conditions \ref{eq:fastCondition}, \ref{eq:slowCondition},
\ref{eq:simpleKe}, and \ref{eq:otherKe} and shown schematically in
Fig. \ref{PHASEDIAGRAM}.

%
%

Our model provides a mechanistic basis for a number of experimental
measurements and observations. For example, the dual entry pathways of
certain viruses suggest that under certain conditions (delineated in
Fig. \ref{PHASEDIAGRAM}) inhibition of fusion does not necessarily
preclude viral entry through endocytosis. HIV fusion inhibitors such
as Enfuvirtide (T-20) bind the intermediary spike-CD4 complex of HIV-1
\cite{T20,T20b}, and reduce $k_{f}$ by preventing CCR5 from inducing
fusion.  Maraviroc binds CCR5 and prevents it from binding the
spike-CD4 complex, effectively reducing $q_{1}$ and also preventing
fusion \cite{EsteTelenti07}.  Since the virus may still enter through
the endocytosis pathway, our analysis suggests that the effectiveness
of fusion inhibitors relies on endocytic entry being less infective
than fusion.

The sensitivity of entry of HIV strains to cell CD4 and CCR5 levels
have recently been quantitatively studied using cells lines in which
expression levels of receptor and coreceptor can be independently
varied \cite{BENHUR}.  This system provides a way of independently
varying $p_{1}$ and $q_{1}$, and has revealed qualitatively different
usage patterns by different HIV strains. Our model provides an
additional dimension to the analysis of receptor/coreceptor tropism.
If endocytic entry {\it does not} significantly diminish the
probability of nuclear entry and productive infection, it is possible
that strains with similar infectivities actually prefer different
entry pathways.

Infection measurements using for example, luciferase reporting of p24 coat
protein levels after productive infection, cannot directly determine
entry pathways. However, using single molecule imaging techniques,
both the timing and entry pathways can be directly observed
\cite{BrandenburgZhuang07,Sieczkarski-02,IMAGING_GERMAN}.  Such direct
imaging techniques may be able to distinguish the mean conditional
times to fusion and endocytosis, particularly in systems with large
fusion and endocytosis rates as shown in Fig. \ref{fig:meanTime}.

Additionally, kinetic studies have been performed to extract the
stoichiometry of receptors and coreceptors per spike, per fusion event
\cite{KABAT0,KABAT1,KABAT2,sodroski1,sodroski2}. Even though our
analysis was based on an intrinsic molecular stoichiometry of one
spike, one receptor, and/or one coreceptor, it implies that the
apparent stoichiometry can vary depending on the degree of wrapping,
and on average, the number of spikes that are
receptor/coreceptor-engaged prior to fusion or endocytosis. The
apparent stoichiometries are defined by $\langle m_{f}\rangle$ and
$\langle n_{f}\rangle$ derived from our model and shown in
Fig. \ref{fig:meanRecept}.  Cells with higher surface densities of
coreceptors, and hence larger $q_{1}$, would more likely fuse before
significant wrapping and formation of spike-receptor-coreceptor
complexes occur. Therefore, a high coreceptor binding rate can present
a {\it lower} apparent coreceptor stoichiometry. It would thus be
interesting to measure kinetics and correlate
spike/receptor/coreceptor stoichiometry across viral strains with
different apparent usage stoichiometries, and across cell types with
varying concentrations of surface receptors and coreceptors.

The assumption that the viral spikes are evenly distributed on the
surface of the virus is valid only if the spikes are immobile on the
virus surface during the entry process.  Freely diffusing glycoprotein
spikes will preferentially bind to membrane receptors or coreceptors
when the spikes come near the cell membrane.  Thus, spikes with
receptors and coreceptors bound would tend to cluster near the bottom
of the virus, precluding full wrapping.  In this case, the probability
that the virus enters the cell via fusion would be increased. It is
also possible that the viral glycoproteins form functional clusters on
the viral envelope \cite{GrunewaldSteven03}.  It is known that the
glycoprotein spikes of recently budded HIV-1 are associated with the
underlying matrix proteins, but that proteolysis occurs during the
maturation process \cite{MATURATION0,MATURATION1}.  If softening of a
maturing virus particle \cite{MATURATION1} also increases glycoprotein spike
mobility, one would expect that mature HIV-1 would be biased towards
using the fusion pathway.

The model we have developed considers only the rudimentary receptor
engagement processes prior to fusion or endocytosis.  Nonetheless,
many more complex mechanisms can be described by our model provided
the effective rate parameters are properly interpreted, or the model
is augmented to include other intermediary processes. For example,
consider the possibility that binding of the virus to a cell surface
receptor activates an endocytic pathway that increases the rate by
which the virus is wrapped by the cell membrane.  The increased
wrapping rate may be the result of, for example, 
a decreased effective stiffness of the cytoskeleton that allows the virus to more
easily enter the cell \cite{MatarreseMalorni05}. An endocytotic
pathway may also rely on the clustering cell surface receptors and/or
coreceptors, as observed in \cite{QiChakraborty01} resulting in a high
local receptor/coreceptor concentration near the virus, thereby {\it
effectively} increasing the rate of receptor, and possibly coreceptor
binding.  

Such a viral entry process can be incorporated within our model by
assuming that prior to activation, receptors and coreceptors bind with
rates $p_1^i$ and $q_1^i$, and that after activation receptors and
coreceptors bind with rates $p_1^a$ and $q_1^a$, respectively.  We
further assume that activation occurs some time $\tau_a$ after the
first receptor binds.  And, for simplicity, we will again consider
that $M=N$ and $r=1$.  In the absence of an active endocytosis
process, two conditions were required for endocytosis to occur: 1) the
virus had to reach the fully wrapped state, and 2) endocytosis had to
be faster than fusion in the fully wrapped state.  If, however, the
cell must initiate an active process for endocytosis, an additional
condition arises: 3) the cell must reach the activated state without
the virus undergoing fusion.  All three conditions must be satisfied
if activated endocytosis is to occur. If activation is important, the
inactivated receptor binding rate $p_1^i$ is slow such that on
average, few receptors bind before activation occurs and
$p_1^i\sqrt{M}\lesssim 1/\tau_a$. In this case, the third condition
can be described in terms of the effective binding rates as follows.


If the inactivated coreceptor binding rate is fast compared with the
time scale on which activation occurs ($q_1^i\gtrsim1/\tau_a$), the
virus will survive to the activated state provided $\tau_a k_f \ll 1$.
If the inactivated coreceptor binding rate is {\it slow} ($q_1^i\ll
1/\tau_a$), it is unlikely that a coreceptor will bind before the
activated state is reached, and the virus will become activated for
any $k_{f}$. The delay time $\tau_{a}$ required to activate the cell's
endocytosis machinery will be relevant if the third condition that the
cell reaches the activated state before viral fusion is not met, but
conditions 1) and 2) are. In this case, a model without the activation
step would predict that the virus should undergo endocytosis when it
in fact will undergo fusion.  In Table 2, we summarize the criteria
under which all three activated endocytosis conditions are met.

We can also consider the case in which cells undergo clathrin or
caveolin-dependent endocytosis that competes with the fusion process
\cite{Sieczkarski-02,VIRALENDOREV,zhuang_flu}.  In these cases, the
membrane adhesion, or wrapping rate $p_{m,n}$ is no longer a function
receptor concentration, but is rather a function of the rate of
assembly of clathrin subunits \cite{CLATHRIN} or calveolin, $M$ of
which cover the virus. This rate would be a function of clathrin or
calveolin concentration, or of molecules that recruit them. If the
formation of clathrin pits or calveolae occurs successively in an
approximately axisymmetric manner, we expect the functional form for
$p_{m,n}$ would be unchanged from Eq. \ref{PN}.  Three variants of our
model could apply to fusion under clathrin or calveolin-mediated
endocytosis. 1) If coreceptors can continue to bind viral spikes and
induce fusion in regions of the membrane that are coated with clathrin
or caveolin, the model described in this work is directly applicable.
In this case, monomeric clathrin/caveolin are ``receptors'' and $M$ is
the total number of monomers required to encapsulate the virus. 2) If
receptor binding, but not fusion, is precluded in regions of the
membrane coated by clathrin/caveolin, the coreceptor binding rate is
no longer given by $q_{m,n} = q_1 (n^{*}(m)-n)$.  In this case,
coreceptors, like monomers of clathrin or caveolin, only bind along
the perimeter of the coated membrane region.  The coreceptor binding
rate then has a form similar to the ``receptor'' (monomeric
clathrin/caveolin) binding rate and is given by $q_{m,n} \approx q_1 N
\sqrt{1-\left(1-\frac{2m}{M}\right)^2}$.  All other aspects of the
model would remain unchanged.  3) If coreceptors within a region of
the membrane coated by clathrin/caveolin cannot induce fusion, the
virus can still undergo fusion if coreceptors bind to spikes along the
perimeter of the coated region {\it and} induce fusion before the
coated region grows enough to cover the location of the coreceptor.
When coreceptor binding is fast compared to the rate at which the
protein scaffold assembles, the instantaneous fusion rate is
proportional to the number of spikes near the contact region.  Instead
of $nk_f$, the effective $m-$dependent fusion rate
$k_fN\sqrt{1-\left(1-\frac{2m}{M}\right)^2}$ arises.  The fusion rate
depends only on the number $m$ of bound receptors and the total number
$N$ of coreceptors. It is independent of the number $n$ of bound
coreceptors, rendering the state space effectively one-dimensional.

\vspace{4mm}

\noindent This work was supported by the NSF (DMS-0349195) and by the
NIH (K25AI058672). SAN also acknowledges support from an NSF Graduate
Research Fellowship. The authors also thank the Institute for Pure and
Applied Mathematics at UCLA for sponsoring a program on Cells and
Materials during which some of this research was conceived and
performed.

\vspace{5mm}
\section{Appendix: Method of Characteristics}

Using specific forms for the receptor and coreceptor attachment rates
$p_{m,n}$ and $q_{m,n}$, analytic expressions for the wrapping and
endocytosis probabilities can be obtained in the large spike number
limit $M\equiv 1/\ve \rightarrow \infty$. Assuming binding rates given
by Eqs.~\ref{PN} and \ref{QN} and defining $x= m\ve$, $y=rn\ve$, and
time $\tau = 2p_{1}t$, we find find the continuum limit of the Master
equation:

\begin{equation}
{\partial P(x,y,\tau) \over \partial \tau} + 
\nabla\cdot\left[{\bf u}(x,y)P(x,y,\tau)\right] = -\kappa_{f}y P(x,y,t).
\label{MASTERCONT}
\end{equation}

\noindent In Eq. \ref{MASTERCONT}, the convection is defined by 

\begin{equation}
{\bf u}(x,y) = (\sqrt{x(1-x)}, \lambda (x-y)),
\end{equation}

\noindent where $\lambda \equiv q_{1}/(2p_{1})$ and $\kappa_{f} =
rk_{f}/(2p_{1}\ve)$ are renormalized coreceptor binding and fusion
rates.
Assuming that the system starts in the state $P(x,y,0) =
\delta(x-\ve)\delta(y)$ (only one receptor attached), the total
derivative of $P(x(\tau),y(\tau),\tau)$ obeys

\begin{equation}
{\dd P(\r(\tau),\tau) \over \dd \tau} = -\kappa_{f}y(\tau)P(\r(\tau),\tau)
\label{TOTAL}
\end{equation}

\noindent provided 

\begin{equation}
{\dd \r(\tau) \over \dd \tau} = \u(x(\tau),y(\tau)).
\label{TRAJECTORY}
\end{equation}

\noindent First consider times before the 
virus is fully wrapped by the cell membrane. 
The components of Eq. \ref{TRAJECTORY} give

\begin{equation}
{\dd x(\tau) \over \dd \tau} = \sqrt{x(\tau)(1-x(\tau))}
\label{TRAJX}
\end{equation}

\noindent and 

\begin{equation}
{\dd y(\tau)\over \dd \tau} = \lambda (x(\tau)-y(\tau)).
\label{TRAJY}
\end{equation}

\noindent Upon using the initial conditions $x(0) = \ve (\approx 0)$ and $y(0) = 0$, 
Eqs. \ref{TRAJX} and \ref{TRAJY} are solved by

\begin{equation}
x(\tau) = {1\over 2}(1-\cos \tau)
\end{equation}

\noindent and 

\begin{equation}
y(\tau) = {1\over 2}-{\lambda^{2}\cos\tau + \lambda\sin\tau + e^{-\lambda\tau} \over
2(\lambda^{2}+1)}\leq x(\tau).
\end{equation}

\noindent Full wrapping of the virus, if it occurs, is defined by 
$x(\tau^{*}) = 1$, where $\tau^{*} = \pi$. Therefore, at time $\tau=\tau^{*}=\pi$, we 
can find the fraction of bound coreceptors as

\begin{equation}
y(\tau^{*}) \equiv y^{*} = {2\lambda^{2}+1-e^{-\lambda\pi}\over 2(\lambda^{2}+1)}<1.
\label{YSTAR}
\end{equation}

Using the forms for the trajectory $\r(\tau)$, 
the probability density for times $\tau \leq \tau^{*}$ can be found
upon solving Eq. \ref{TOTAL} to give

\begin{equation}
\ln P(\tau) = \kappa_{f}\left[{1\over 2\lambda}-{\tau \over 2} +
{\lambda^{2}(\lambda\sin\tau - \cos\tau) - e^{-\lambda\tau}\over 2\lambda(\lambda^{2}+1)}\right].
\label{LNPTAU}
\end{equation}

\noindent The probability density $P^{*}$ that the virus reaches the
fully wrapped state (the continuum analogue of $P_{M}$ shown in
Fig. \ref{fig:nonDim}) is found by evaluating
$P(x(\tau^{*})=1,y(\tau^{*})=y^{*},\tau^{*})\equiv P^{*}$.  This
evaluation gives Eq. \ref{eq:Pfull} in the large $M,N$ limit.

At times $\tau > \tau^{*}$, additional receptors cannot bind, thus,
$x(\tau > \tau^{*}) = 1$, and $y(\tau)$ follows

\begin{equation}
{\dd y(\tau) \over \dd\tau} = \lambda (1-y(\tau)).
\label{dydt}
\end{equation}

\noindent Upon defining $z^{*}\equiv 1-y^{*}$, 
Eq. \ref{dydt} is solved by 

\begin{equation}
y(\tau > \tau^{*}) = 1-z^{*}e^{-\lambda(\tau-\pi)}.
\end{equation}

\noindent In terms of the renormalized endocytosis rate $\kappa_{e} =
k_{e}/(2p_{1})$, the probability that the virus has not entered the
cell through either fusion or endocytosis at time $\tau$ follows

\begin{equation}
{\dd P(x(\tau),y(\tau),\tau)\over \dd \tau} = 
-(\kappa_{f}+\kappa_{e})P(x(\tau),y(\tau),\tau), \quad \tau > \tau^{*}
\end{equation}

\noindent which is solved by

\begin{equation}
\begin{array}{ll}
P(\tau) = P^{*}\exp\bigg[-(\kappa_{f}+\kappa_{e})(\tau-\tau^{*}) \\[13pt]
\:\hspace{2.8cm} \displaystyle +{\kappa_{f}\over \lambda}z^{*}(1-e^{-\lambda(\tau-\tau^{*})})\bigg].
\end{array}
\end{equation}

\noindent The probability $Q_{e}$ that the virus particle undergoes endocytosis is then given by

\begin{equation}
Q_{e} = \displaystyle \kappa_{e}\int_{\pi}^{\infty}P(\tau)\dd\tau =  {\kappa_{e}\over \lambda}P^{*}
e^{\kappa_{f}z^{*}/\lambda}\left({\kappa_{f}\over \lambda}z^{*}\right)^{-(\kappa_{f}+\kappa_{e})/\lambda}
\gamma\left({\kappa_{f}+\kappa_{e}\over \lambda},{\kappa_{f}z^{*}\over \lambda}\right).
\label{eq:ContinQe}
\end{equation}

\noindent where $\gamma$ is the incomplete lower Gamma function.  This
expression was used to generate the analytic results plotted in
Fig. \ref{fig:compare}. 

Besides our results obtained using the specific forms of receptor and
coreceptor binding rates, conditions analogous to those in
Eqs. \ref{eq:fastCondition} and \ref{eq:slowCondition} can also be
obtain for general coreceptor-independent binding rate $p_{m}$ by
using simple scaling arguments.  When coreceptor binding is fast,
$q_{1} \gtrsim p_{1}$, the probability of fusion is approximately
$k_{f} \langle n_{f}\rangle t^{*}$, where

\begin{equation}
t^{*} \approx \sum_{m=1}^{M}{1 \over p_{m}}
\end{equation}
is the mean conditional wrapping time, and 
$\langle n_{f}\rangle$ is the mean number of 
bound coreceptors before fusion. For fast 
coreceptor binding $\langle n_{f}\rangle \sim N$, and 
the necessary (but not sufficient) condition for 
full virus wrapping ($P_{M} \sim 1$) is

\begin{equation}
k_{f} N \sum_{m=1}^{M}{1\over p_{m}} = rk_{f}M\sum_{m=1}^{M}{1\over p_{m}} \ll 1.
\label{GENCONDFAST}
\end{equation}
When coreceptor binding is slow, $\langle n_{f}\rangle 
\approx q_{1} N t^{*}$ increases linearly with both time and the 
number of  available coreceptor-biding spikes. In this case, the 
necessary condition for virus wrapping becomes

\begin{equation}
k_{f}q_{1}N(t^{*})^{2} = rk_{f} M q_{1}\left(\sum_{m=1}^{M}{1\over p_{m}}\right)^{2} \ll 1.
\label{GENCONDSLOW}
\end{equation}
Upon inserting the smoothly varying forms for $p_{m}$ from
Eq.~\ref{PN} into the above relationships, they reduce to the
conditions \ref{eq:fastCondition} and \ref{eq:slowCondition},
respectively.

\vspace{4mm}

\noindent This work was supported by the National Science Foundation
through grant DMS-0349195, and by the National Institutes of Health
through grant K25AI41935.  SAN was also supported by a National
Science Foundation Graduate Research Fellowship.

\bibliography{two9.bbl}

\newpage

\section*{Tables}

\vspace{5mm}
\begin{tabular}{|l|l|l|}
\hline
Quantity & HIV-1 & HSV-1 \\
\hline
\hline
radius $R$  & 0.05$\mu$m  \cite{MahyCompans96}  &0.1$\mu$m \cite{WilsonSande01} \\ \hline
\multirow{2}{*}{spikes/virus} &  &235-480 gD \cite{ClarkeKlenerman07} \\
\cline{3-3} & 8-14 \cite{TRIMERS,SPIKESEM} & $\sim$700 total \cite{GrunewaldSteven03}\\ \hline
receptor & $K_{D} \approx 5nM$ \cite{BINDINGE} & \\
\parbox{2pt}{\hspace{2pt}}binding  & $\Delta H \approx -100 k_B T$ 
\cite{BINDINGE}  & \\\hline
coreceptor  & $K_{D} \approx 4nM$ \cite{GERARD} &  \\ 
\parbox{2pt}{\hspace{2pt}}binding & $\Delta H \approx -300k_{B}T$ 
\cite{CCR5} &\\\hline
receptor diff. &  & \\
\parbox{12pt}{\hspace{12pt}}const. $D_r$ & \multirow{2}{*}{$0.044 \mu\rm{m}^2/s$ \cite{Finnegan-07}}&\\\hline
coreceptor diff.&   & \\
\parbox{12pt}{\hspace{12pt}}const. $D_c$&\multirow{2}{*}
{$0.05 \mu\rm{m}^2/s$  \cite{Finnegan-07}} & \\ \hline
\multirow{2}{*}{host cell radius} & T-cell & Epithelial cell  \\
&$4\mu\rm{m}$\cite{ShenLouie99}&$5 \mu\rm{m}$\cite{Shaw96}\\ \hline
cell receptor & 300-3000 & $6-9\times 10^6$\\
\parbox{12pt}{\hspace{12pt}}density &\parbox{12pt}{\hspace{12pt}}CD4/$\mu$m$^2$
\cite{Platt-98}&\parbox{12pt}{\hspace{12pt}}HS/$\mu$m$^2$\cite{BaumannKeller97}\\\hline
cell coreceptor  &&\\
\parbox{12pt}{\hspace{12pt}}density &\multirow{2}{*}{60 CCR5$/\mu$m$^2$ \cite{Platt-98}}& \\  \hline
\end{tabular}

\vspace{7mm}

\noindent Table 1: Known representative parameter values for virus
spikes, receptors, and coreceptors

\vspace{1cm}


\vspace{5mm}
\begin{tabular}{|c|c|c|c|c|c|c|}
\hline && Condition 1: && Condition 2: && Condition 3: \\ 
& fast or & survival to & fast or &endocytosis &  fast or &  survival to  \\
& slow &fully wrapped & slow & faster than & slow & activated \\
& def. & state & def. & fusion & def. & state \\ \hline fast
coreceptor &\multirow{2}{*}{$\frac{q^a_1}{p^a_1}\gtrsim
1$} &\multirow{2}{*}{$\frac{k_fM}{p_1^a}\ll 1$}
&\multirow{2}{*}{$\frac{q^a_1}{p^a_1}\gtrsim 1$}
&\multirow{2}{*}{$k_e\gg Nk_f$} & \multirow{2}{*}{$q^i_1\tau_a\gtrsim 1$}
&\multirow{2}{*}{$\tau_ak_f\ll1$}\\ binding &&&&&&\\ \hline slow
coreceptor 
&\multirow{2}{*}{$\frac{1}{N}\!\ll\!\frac{q_1^a}{p_1^a}\!\ll\!1$}
&\multirow{2}{*}{$\left(\frac{k_f
M}{p_1^a}\right)\!\left(\frac{q_1^a}{p_1^a}\right)\!\ll\!1$} &
$\frac{q_1^a}{p_1^a} \ll 1,$  &  \multirow{2}{*}{$k_e\gg Nq_1^a$} & \multirow{2}{*}{$q_1^i\tau_a\ll 1$}
&\multirow{2}{*}{$\tau_a^2k_fq_1^i\ll 1$} \\
binding & & &$q_1^aN\ll k_f$ & &  &\\ \hline
\end{tabular}

\vspace{7mm}

\noindent Table 2: Conditions for 1) survival
to fully wrapped state after activation, 2) endocytosis 
being faster than fusion, and 3) reaching the activated state before 
fusion.

\newpage

\section*{Figure Captions}

\noindent Figure 1: A schematic of the kinetic steps involved in receptor and
coreceptor engagement, which ultimately lead to membrane fusion or
endocytosis.  Receptors and coreceptors in the cell membrane are
represented by black line segments and red zig-zags, respectively. The
projected contact area nucleated by the number of bound receptors is
also shown.  Only viral spikes that have a coreceptor bound can induce
fusion.  Endocytosis can occur only when the contact region grows to
the surface area of the virus particle.  Left: The receptors and
coreceptors both bind to the same viral spikes (blue circles).  An
example of such as virus is HIV-1, where spikes, likely composed of
trimers of gp120/41, bind to both CD4 and CCR5.  Right: An example
(such as Herpes Simplex Virus) in which coreceptors and receptors bind
to different spikes, with the ratio of receptor-binding spikes (blue
circles) to coreceptor-binding spikes (yellow hexagons) defined by
$r$.

\vspace{1cm}

\noindent Figure 2: Two-dimensional state space for receptor and
coreceptor-mediated viral entry.  Each state corresponds to a virus
particle bound to $m\leq M$ receptors and $n\leq N=rM$ coreceptors. In
this example, the fraction of coreceptor-binding spikes to
receptor-binding spikes is $r=1/2$. The probability fluxes through the
fusion and endocytosis pathways are indicated by the red and green
arrows, respectively. A representative trajectory of the stochastic
process that results in endocytosis is indicated by the blue dashed
curve.

\vspace{1cm}

\noindent Figure 3: A schematic of a partially wrapped virus
particle. The unbound spikes above the contact region are represented
by light blue circles, while the receptor-bound spikes in the contact
region are represented by the dark blue circles. Spikes that are bound
to coreceptors are indicated by the red-filled circles. The unbound
receptors and coreceptors on the cell membrane (green) are not shown.

\vspace{1cm}

\noindent Figure 4: (a) The probability that the virus undergoes
endocytosis is plotted as a function of the normalized fusion rate,
$k_{f}/p_{1}$, for different values of the normalized coreceptor
binding rate, $q_{1}/p_{1}$.  The probability of endocytosis decreases
with increasing fusion rate and, for a given fusion rate, the
probability of endocytosis increases with decreasing $q_{1}/p_{1}$.
In this example, the normalized endocytosis rate, $k_e/p_{1}=1$.  (b)
For $q_{1}/p_{1}=1$, the probability of endocytosis is plotted as a
function of fusion rate for different values of the normalized
endocytosis rate, $k_e/p_{1}$.  In both plots, the number of
receptor-binding spikes and the number of coreceptor-binding spikes
are set to $M=N=100$.

\vspace{1cm}

\noindent Figure 5: Endocytosis rates are plotted as a function of
$M=N$.  During wrapping, the fusion rate is proportional to the number
of bound coreceptors, and increases with increasing $N$ (in this case
equal to $M$).  The probability that the virus enters the cell through
endocytosis decreases with increasing $M=N$.

\newpage

\noindent Figure 6: The exact numeric solution of Eqs. \ref{eq:Master}
and \ref{eq:endoFlux} for the probability $Q_e$ that the virus
undergoes endocytosis is plotted as a function of $\k_f\equiv rk_f M
/(2p_1)$, the dimensionless fusion rate and compared to the
$M\rightarrow \infty$ asymptotic solution (thin solid curves).  Two
sets of curves, corresponding to $\lambda \equiv q_1/(2p_1)=0.1, 2$
are shown for $M=N =10,100$, and $1000$ ($r=1$). In these plots, the
endocytosis rate was taken to be $k_e/p_1=2$.

\vspace{1cm}

\noindent Figure 7: Wrapping probabilities for $M=N$ ($r=1$).  (a) For
$q_{1}\gtrsim p_{1}$, the probability $P_M$ that the virus reaches the
fully wrapped state is plotted as a function of the dimensionless
fusion rate parameter $rk_f M/p_{1}$.  When this parameter is small,
$P_M$ approaches unity, but when $rk_f M/p_{1} \gg 1$, $P_M$ is
small. (b) When $1/N \ll q_{1}/p_{1}\ll 1$, the wrapping probability
$P_M$ is plotted as a function of the dimensionless expression $(rk_f
M/p_{1})(q_1/p_{1})$.  In this case the transition of $P_{M}$ from
large to small values occurs at $(rk_f M/p_{1})(q_1/p_{1}) \sim O(1)$
In both plots, only one parameter was varied within a group of symbols
of the same color and shape.  The number of spikes $M$ was varied
within the groups of circles, and the fusion rate, $k_f$ was varied
within the groups of triangles.

\vspace{1cm}

\noindent Figure 8: (a) Normalized mean times to fusion and
endocytosis plotted as functions of $k_f/p_1$, the fusion rate per
coreceptor-spike complex.  Parameters used were $M=N= 100$,
$q_{1}/p_{1} = 50$, $k_e/p_1 = 0.001$. (b) Normalized mean times to
fusion and endocytosis plotted as functions of $k_e/p_1$.  Here, $M=N=
100$, $q_{1}/p_1 = 5$, and $k_f/p_1 = 10^{-6}$, were used.  For
reference, $Q_e$, the corresponding probability that the virus
undergoes endocytosis is also plotted.

\vspace{1cm}

\noindent Figure 9: (a) The mean number of receptors bound at the
moment of viral fusion, and the mean number of receptors bound at the
moment of viral entry (regardless of entry pathway) plotted as
functions of the normalized coreceptor binding rate, $q_{1}/p_{1}$.
(b) The mean number of coreceptors bound at the moment of fusion and
endocytosis, and the average number of coreceptors bound are plotted.
The probability that the virus undergoes endocytosis, $Q_e$ is plotted
for reference.  For both plots $M=N=100$, $k_f/p_1 = 0.1$, $k_e/p_1 =
1$.

\vspace{1cm}

\noindent Figure 10: Qualitative ``phase diagram'' showing the regimes
of parameter space in which endocytosis is dominant. Diagrams (a),(b),
and (c), correspond to fast, intermediate, and slow coreceptor
binding, respectively. In all diagrams, parameters falling within the
blue region left of the vertical thick dashed line favor full viral
wrapping before fusion occurs ($P_{M}\approx 1$).  In the yellow
sector above the thin dashed curves, the rate of endocytosis exceeds
the effective rate of fusion in the fully wrapped state.  In the green
intersection of these regions, the virus is likely to reach the fully
wrapped state {\it and} undergo endocytosis.  Note that when
coreceptor binding is very slow (c), the virus reaches the fully
wrapped state for all values of $k_f$.

\newpage

\section*{Figures}

\vspace{1cm}

\begin{figure}[ht]
\begin{center}
\includegraphics[width=3.2in]{Fig4.eps}
\end{center}
\caption{}
\label{fig:PeVkf}
\end{figure}

\begin{figure}[ht]
\begin{center}
\includegraphics[width = 3.2in]{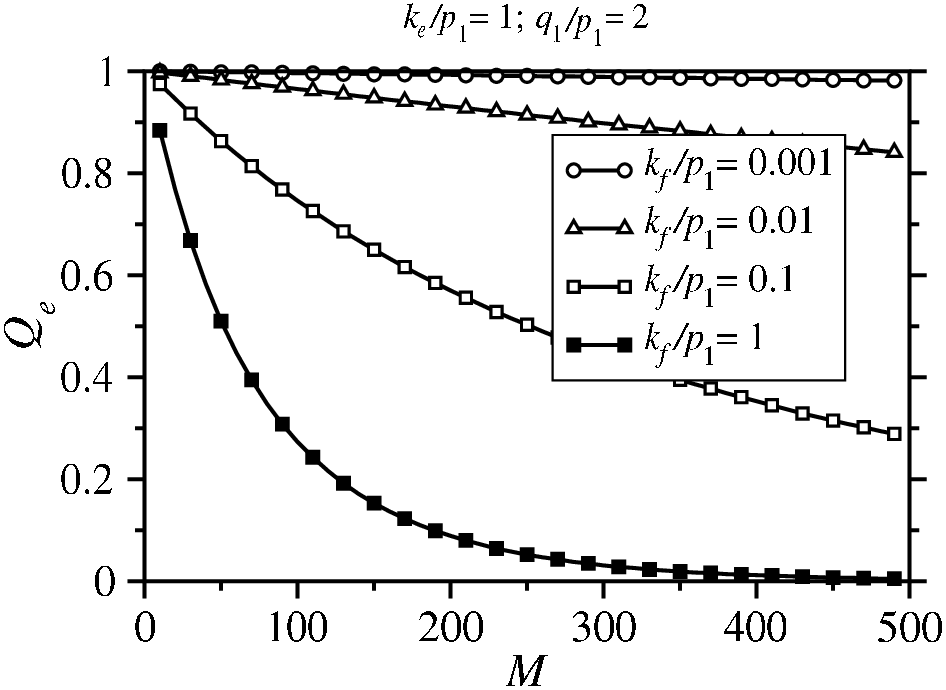}
\end{center}
\caption{}
\label{fig:peVm}
\end{figure}

\vspace{2cm}

\begin{figure}[ht]
\begin{center}
\includegraphics[width = 3.2in]{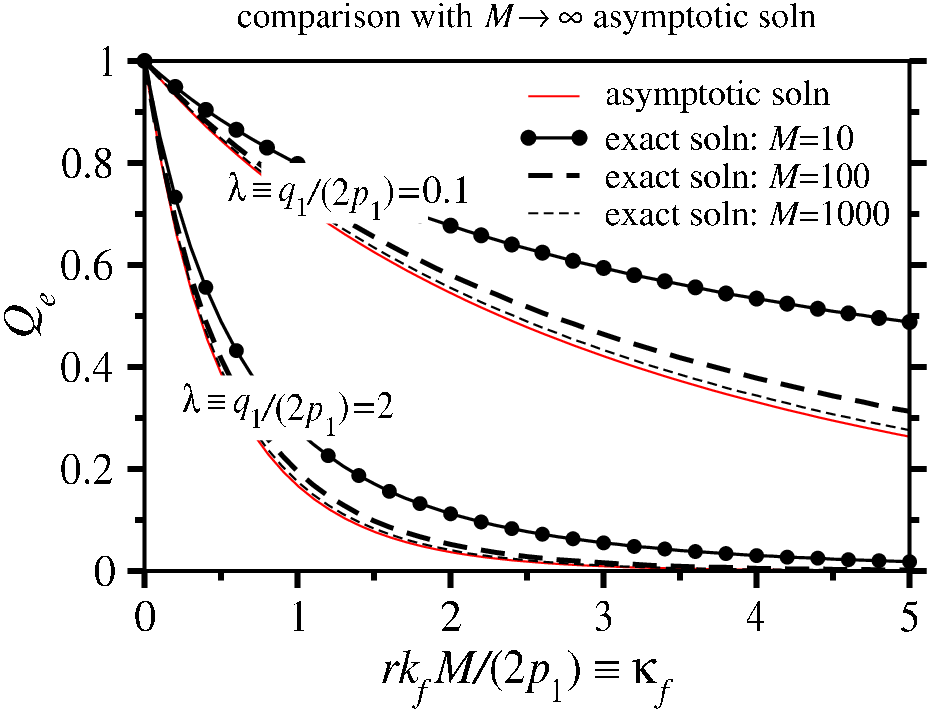}
\end{center}
\caption{}
\label{fig:compare}
\end{figure}

\begin{figure}[ht]
\begin{center}
\includegraphics[width = 3.2in]{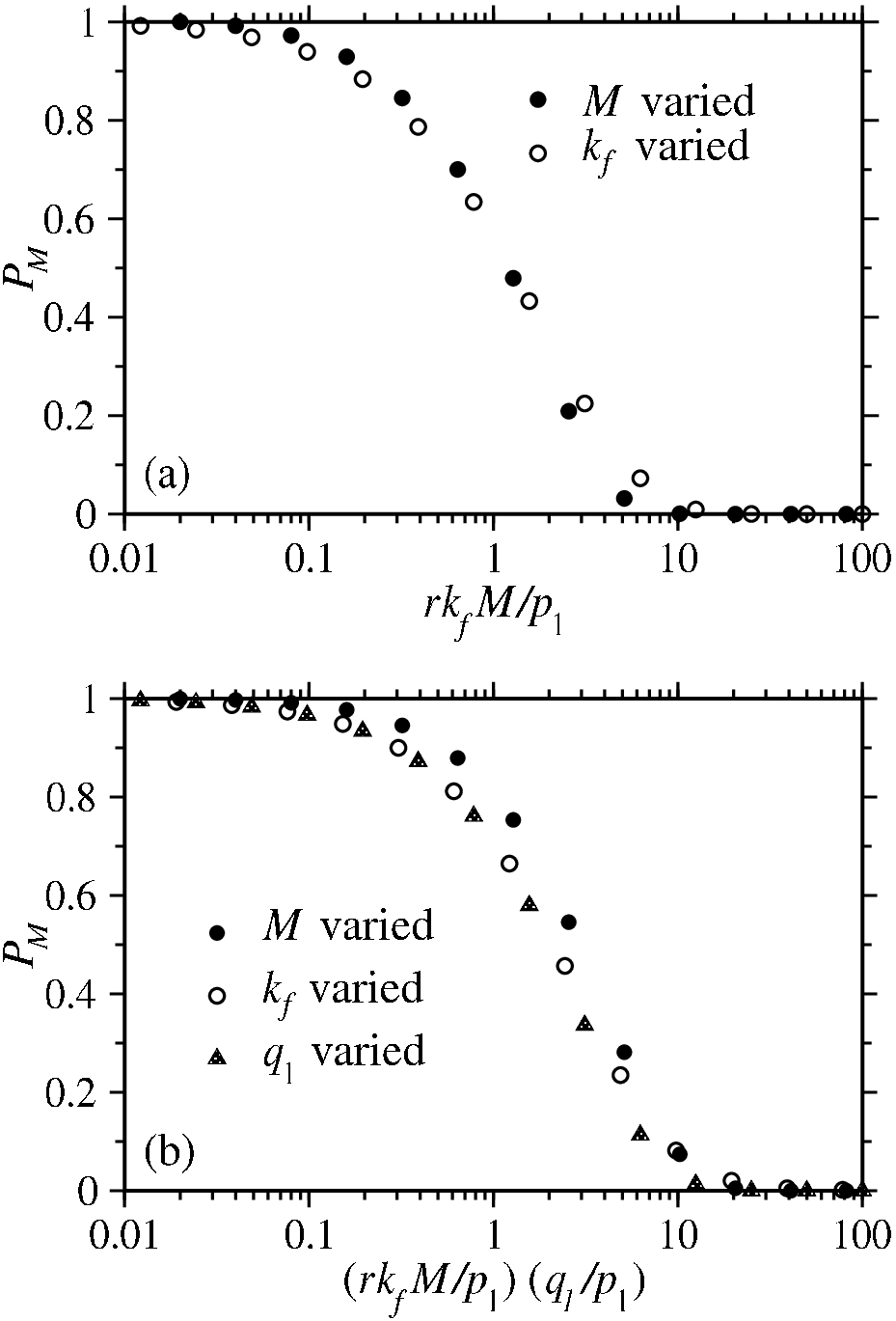}
\end{center}
\caption{}
\label{fig:nonDim}
\end{figure}

\begin{figure}[ht]
\begin{center}
\includegraphics[width = 3.2in]{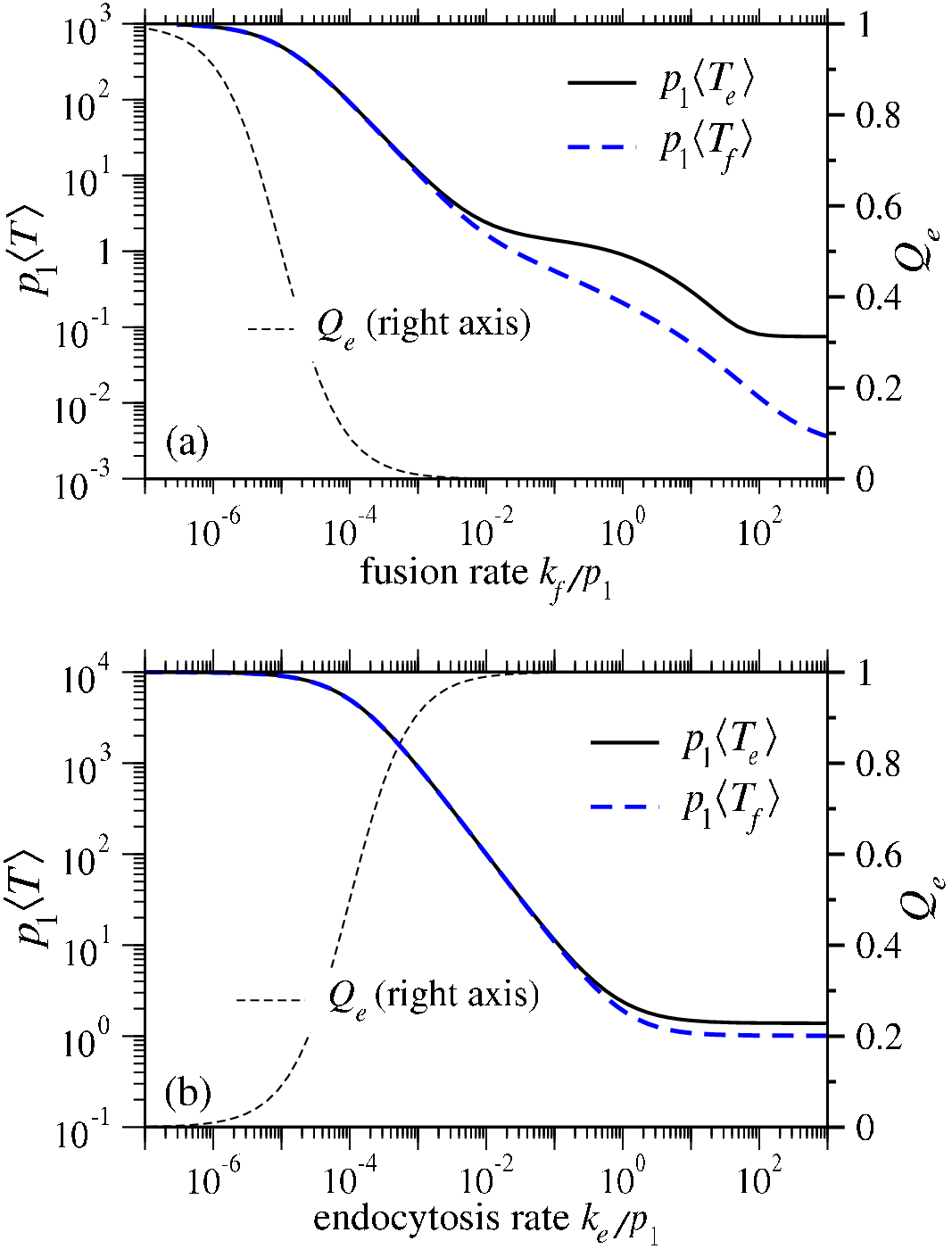}
\end{center}
\caption{}
\label{fig:meanTime}
\end{figure}

\begin{figure}[ht]
\begin{center}
\includegraphics[width = 3.2in]{Fig9.eps}
\end{center}
\caption{}
\label{fig:meanRecept}
\end{figure}

\begin{figure}[ht]
\begin{center}
\includegraphics[width=6.4in]{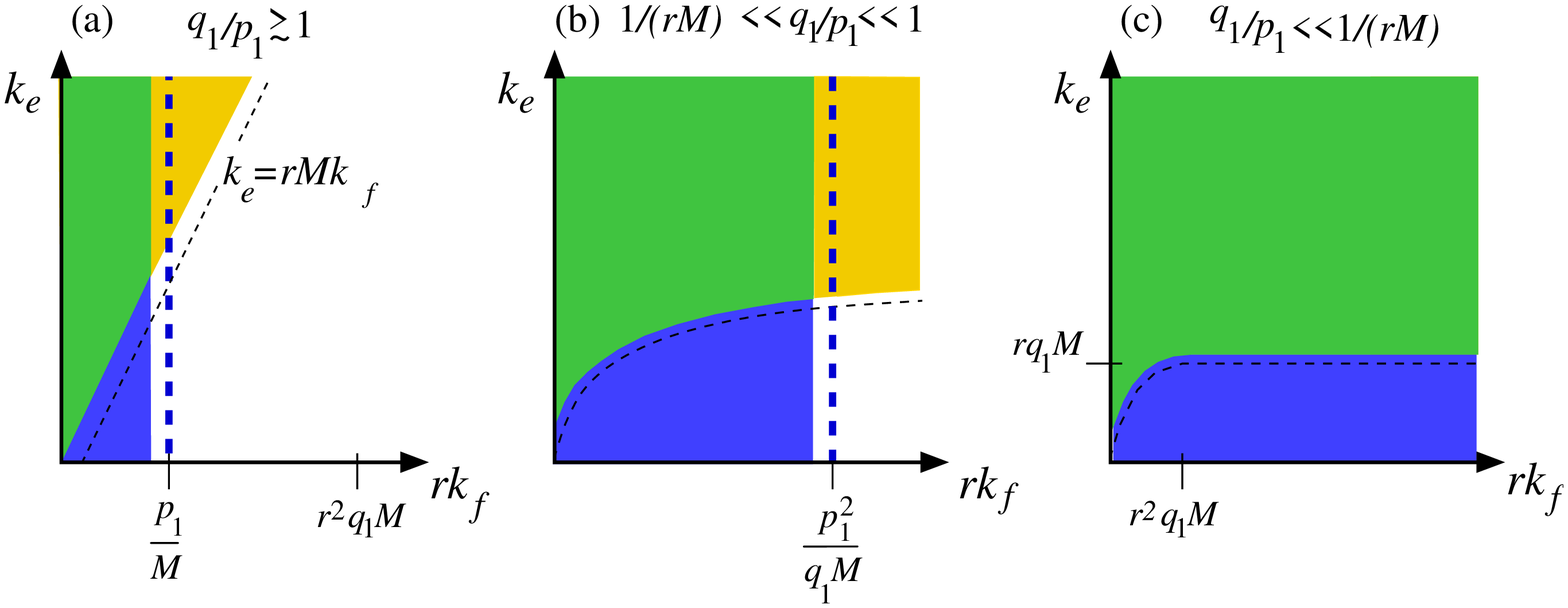}
\end{center}
\caption{}
\label{PHASEDIAGRAM}
\end{figure}

\end{document}